\documentclass[prd,aps,amsmath,amssymb,preprintnumbers,twocolumn,showpacs,floats]{revtex4}
\input{epsf}
\usepackage{graphics}
\usepackage{amsmath}
\usepackage{amsfonts}
\usepackage{amssymb}
\begin{document}
\title{Spin operator and spin states in Galilean covariant Fermi field theories}
\author{Fuad M. Saradzhev}
\affiliation{Centre for Science, Athabasca University, Athabasca,
Alberta, Canada\footnote{\tt E-mail address: fuads@athabascau.ca}}

\begin{abstract}

Spin degrees of freedom of the Galilean covariant Dirac field in $(4+1)$ dimensions and its non-relativistic counterpart in $(3+1)$ dimensions are examined. Two standard choices of spin operator, the Galilean covariant and Dirac spin operators, are considered. It is shown that the Dirac spin of the Galilean covariant Dirac field in $(4+1)$ dimensions is not conserved, and the role of non-Galilean boosts in its non-conservation is stressed out. After reduction to $(3+1)$ dimensions the Dirac field turns into a nonrelativistic Fermi field with a conserved Dirac spin. A generalized form of the L\'{e}vy-Leblond equations for the Fermi field is given. One-particle spin states are constructed. A particle-antiparticle system is discussed.

\end{abstract}


\maketitle

\section{Introduction}

Galilean invariance underlies many low-energy systems encountered in nuclear physics, condensed matter physics and many-body theory. It is used in the field theoretical study of superfluids,
superconductors, and Bose-Einstein condensation \cite{chai}-\cite{grei}.

A $(4+1)$-dimensional covariant formulation of Galilean covariance \cite{khan} based on an extended space-time approach developed in \cite{taka},\cite{omo} provides new insights into the properties of these systems.
A formulation of Galilean covariance within a relativistic framework in one
higher dimension makes non-relativistic theories similar to Lorentz covariant
theories. Many procedures and calculations can be carried out in the same way as
relativistic ones. Some characteristic features of non-relativistic theories
are likely to appear only after a reduction to $(3+1)$ dimensions.
Historically, a dimensional reduction from $(4+1)$ to $(3+1)$ dimensions was first developed for Lorentz covariant theories \cite{pin}. The extended space-time formulation allows us to treat relativistic and
non-relativistic theories in $(3+1)$ dimensions in a unified approach, depending
on how $(3+1)$-dimensional space-time is embedded into a $(4+1)$-dimensional
manifold. A similar approach was used in the investigation of fluid dynamics in
\cite{jack},\cite{has}.

In the present paper, we study spin degrees of freedom of nonrelativistic Fermi fields
in both $(4+1)$ and $(3+1)$ dimensions. We start with a Galilean covariant Fermi
field theory in $(4+1)$ dimensions and then perform a reduction to $(3+1)$ dimensions.
We want to determine to what extent spin and its characteristics are affected by this
reduction.

In relativistic theories, the orbital angular momentum and Dirac spin of a moving particle are not separately conserved. The component of the Dirac spin in any fixed direction cannot be therefore used to enumerate the particle's spin states. Such enumeration is possible in the nonrelativistic case. Our aim is to demonstrate the Dirac spin conservation in $(3+1)$ dimensions as a result of dimensional reduction and to construct the spin states explicitly.

The concept of spin for nonrelativistic particles was introduced
in \cite{leblond} through the theory of unitary irreducible
representations of the Galilei group \cite{ino}-\cite{levy2}. A nonrelativistic
particle described by such representation is localizable for any
value of spin, if it has a nonzero mass. Although the concept of spin
appears here in the same way as in the relativistic case, the
nonrelativistic spin particles are not involved in spin-orbit interactions
and do not possess electromagnetic multiple momenta \cite{leblond}.

Our paper is organized as follows. In Sect. II, we first briefly review a covariant
formulation of Galilean covariance in $(4+1)$ dimensions. Then we construct the Galilean
covariant spin operator. The expression for this operator was previously given in \cite{group}-\cite{pauri}, and a possible connection between spin and statistics was discussed in
\cite{messiah}. We derive the same expression in a different way similar to the Lorentz covariant
procedure presented in~\cite{bogo}.

We define the five-dimensional Galilei invariant Dirac field model and determine the transformation
properties of the Dirac field. Conserved currents and the corresponding generators are constructed. We observe that the model is invariant with respect to non-Galilean transformations as well and introduce non-Galilean boosts. We discuss similarities and differences between the Galilean covariant and Dirac spin operators and demonstrate the role of non-Galilean boosts in the non-conservation of the Dirac spin. Different choices of spin operator in the relativistic case were studied in \cite{terno}.

In Sect. III, we perform a reduction to $(3+1)$ dimensions by eliminating the $x^5$ coordinate and keeping the model invariant under reduced Galilean transformations. We obtain the effective Lagrangian density for a nonrelativistic Fermi field  in $(3+1)$ dimensions. The Fermi field is quantized, and one-particle states are constructed. We prove that the one-particle states are spin-$1/2$ and use the third component of spin to enumerate them.

Finally, we study a particle-antiparticle system. The possibility to include antiparticles within a Galilean framework was discussed in \cite{levy},\cite{hor}. We show that particles and antiparticles belong to two different mass sectors and construct transition operators that connect states with different number of particles and antiparticles. We conclude with Discussion in Sect. IV.


\section{Galilean covariant Fermi field theories in $(4+1)$ dimensions}


The Galilean covariant theories in $(4+1)$ dimensions are built with
Lorentz-like action functionals, except that Galilean kinematics is based
on the so-called Galilean five-vectors \cite{khan}
\[
({\bf x},x^4,x^5) = \Big({\bf x},\bar{c}t,\frac{s}{\bar{c}}\Big),
\]
where $\bar{c}$ is a parameter with the dimensions of velocity, which will be specified below, and $s$ stands for the additional fifth coordinate. By rescaling
\[
t \to x^4= \bar{c} t, \qquad s \to x^5 = \frac{s}{\bar{c}},
\]
the physical units of space and time are adjusted in such way that all the components of the Galilean five-vectors have the same dimensions.

Under homogeneous Galilean transformations, these vectors transform as follows:
\begin{eqnarray}
{\bf x'} &=& {\bf Rx} - {\boldsymbol{\beta}} x^4,\nonumber\\
{x'}^{4}&=&x^4,
\label{strans} \\
{x'}^{5}&=&x^5-({\bf Rx}) \cdot {\boldsymbol{\beta}} +\frac{1}{2}{|{\boldsymbol{\beta}}|}^{2} x^4, \nonumber
\end{eqnarray}
where $\boldsymbol{\beta} = ({\bf v}/\bar{c})$, ${\bf v}$ is the relative velocity
between two reference frames, ${\bf R}$ is a three-by-three orthogonal rotation matrix.

These transformations leave invariant the scalar product,
$g_{\mu\nu}x^{\mu}x^{\nu}$, ${\mu},{\nu}=1,...,5$,
defined with the Galilean metric
\begin{equation}
g_{\mu\nu}=\left( \begin{array}{ccc} -{\mathbf 1}_{3\times 3}
& 0 & 0 \\ 0 & 0 & 1 \\ 0 & 1 & 0\end{array} \right).\label{galileanmetric}
\end{equation}

The momentum five-vector is
\[
p_{\mu} = {\rm i}{\partial}_{\mu} = \Big({\rm i}\boldsymbol{\nabla}, {\rm i}\frac{{\partial}_t}{\bar{c}},
{\rm i}\bar{c}{\partial}_s\Big) = ({\bf p},p_4,p_5).
\]
With the usual identification ${\rm i}{\partial}_t \to E$, we have $p_4=~E/{\bar{c}}$. Since ${p_5}/{\bar{c}}$ has the dimension of mass, it is identified with the inertial mass $m$ so that the coordinate $x^5$ is canonically conjugate to $p_5 = m\bar{c}$.

The inhomogeneous Galilei group also includes translations in space, time and
$x^5$-direction, being a subgroup of the Poincare group. While the Poincare group
in $(4+1)$-dimensions is generated by 15 elements, five translations $P^{\mu}$ and
ten rotations $M^{{\mu}{\nu}}$ with ${\mu}<{\nu}$, its Galilei subgroup
has 11 generators.

The non-zero commutation relations of the Galilei Lie algebra are
\begin{eqnarray}
{[J^a, J^b]}_{-}= {\rm i} {\epsilon}^{abc} J^c, &  \qquad & {[J^a, K^b]}_{-}={\rm i} {\epsilon}^{abc} K^c, \nonumber\\
{[J^a, P^b]}_{-}= {\rm i} {\epsilon}^{abc} P^c, &  \qquad & {[K^a, P_4]}_{-}={\rm i} P^a,
\label{commutation} \\
{[P^a, K^b]}_{-}= {\rm i} g^{ab} P_5, & \qquad & \nonumber
\end{eqnarray}
where $a,b,c=1,2,3$, $P^a$, $J^a=(1/2){\epsilon}^{abc}M^{bc}$ and
$K^{a}=M^{a4}$ are generators of space translations, rotations and Galilean
boosts, respectively, while $P_4$ and $P_5$ are generators of time and
$x^5$-translations. The explicit expressions for these generators in the case
of Fermi fields will be given below.

Like the Poincare group in $(4+1)$ dimensions, the Galilei group admits
three Casimir invariants:
\begin{eqnarray}
I_1 & = & P_{\mu} P^{\mu}, \nonumber \\
I_2 & = & P_5, \\
I_3 & = & W_{5{\mu}} {W_5}^{\mu}, \nonumber
\end{eqnarray}
where
\[
W_{\mu\nu} \equiv \dfrac{1}{2} {\epsilon}_{\mu\alpha\beta\rho\nu} P^{\alpha} M^{\beta\rho}
\]
is the five-dimensional Pauli-Lubanski tensor, ${\epsilon}_{\mu\nu\alpha\beta\rho}$ being the
totally antisymmetric tensor in five dimensions. We assume that ${\epsilon}_{54abc}=
{\epsilon}_{abc}$.

The invariant $I_2$ is the inertial mass operator, which is assumed to be
positive (or negative) definite. The inertial mass is conserved independently;
that is a characteristic feature of Galilean covariant theories. This implies the so-called
Bargmann superselection rule \cite{barg}, the theory factoring into sectors labeled by the eigenvalue of $I_2$.

The invariant $I_1$ is related to the rest (or internal) energy:
\[
I_1 = 2P_5P_4 - {\bf P}^2 = 2P_5\Big(P_4 - \dfrac{{\bf P}^2}{2P_5}\Big) = 2P_5P_4^{(0)},
\]
where
\begin{equation}
P_4^{(0)} \equiv P_4 - \dfrac{{\bf P}^2}{2P_5}
\label{restenergy}
\end{equation}
is the rest energy operator. In Galilean theories the inertial and rest masses are not
in general the same \cite{jackiw1}. They are taken equal if the Galilean theory
is considered to be the limit of a Lorentz covariant one.

For a state with momentum $p_{\mu}$,
\[
P_{\mu} |p \rangle = p_{\mu} |p \rangle,
\]
assuming that
\begin{equation}
p_{\mu} p^{\mu} = k^2 > 0,
\label{defk}
\end{equation}
this gives us the dispersion relation
\begin{equation}
E |p \rangle = \Big(\frac{{\bf p}^2}{2m} + \frac{k^2}{2m}\Big) |p \rangle.
\label{dispersion}
\end{equation}
The rest energy $k^2/{2m}$ takes the familiar form $m_0 {\bar{c}}^2$ if $\bar{c}$ is defined as
\begin{equation}
\bar{c} = \frac{k}{\sqrt{2 m m_0}},
\label{constc}
\end{equation}
where $m_0$ is the rest mass. Both $m_0$ and $\bar{c}$ are constants with respect to the Galilean transformations.

\subsection{Spin operator}

The axial vector $W_{5{\mu}}$ obeys the following relations
\begin{equation}
P^{\mu} W_{5{\mu}} = 0,
\label{relation1}
\end{equation}
\begin{equation}
{[P^{\sigma}, W_{5{\mu}}]}_{-} = 0,
\label{relation2}
\end{equation}
\begin{equation}
{[W_{5{\mu}}, W_{5{\nu}}]}_{-} = {\rm i}{\epsilon}_{5\mu\nu\lambda\rho} {W_5}^{\lambda} P^{\rho}.
\label{relation3}
\end{equation}
Since $W_{55}=0$, $W_{5{\mu}}$ has four non-vanishing components
\[
W_{5{\mu}} = ( W_{5a}, W_{54}, 0),
\]
where
\begin{eqnarray}
W_{5a} & = & P_5 J_a - {\epsilon}_{abc} K_b P_c,\\
W_{54} & = & P_a J_a.
\end{eqnarray}
As a consequence of Bargmann's superselection rule, the Hilbert space is decomposed into a direct sum
of subspaces representing different inertial mass sectors including a zero-mass one. Let us consider
a non-zero inertial mass subspace. In such subspace the operator $P_5$ is either positive or negative definite
and therefore $P_5^{-1}$ is well defined. As in Lorentz covariant theories
\cite{bogo},\cite{gursey} we define the spin operator components $S^a$ as linear combinations of the components of $W_{5{\mu}}$,
\begin{equation}
S^a = A [{W_5}^a - B W_{54} P^a],
\label{spincomponents}
\end{equation}
where the coefficients $A$ and $B$ depend only on the five-momentum $P_{\mu}$. We find
$A$ and $B$ from the requirement that the usual commutation relations for the spin
operator hold:
\begin{equation}
{[S^a, S^b]}_{-}= {\rm i} {\epsilon}^{abc} S^c.
\label{spincommutation}
\end{equation}
Substituting (\ref{spincomponents}) into (\ref{spincommutation}), this gives us a system
of two equations,
\begin{eqnarray}
A^2 (P_5 - B {\bf P}^2) & = & A, \nonumber \\
A^2 B P_5 & = & - A B. \nonumber
\end{eqnarray}
For nonzero $B$, the system is solved by
\begin{equation}
A = - P_5^{-1}, \qquad B = \frac{2P_5}{{\bf P}^2},
\label{solution1}
\end{equation}
while for $B=0$ we get
\begin{equation}
A = P_5^{-1}.
\label{solution2}
\end{equation}
The first solution is singular in the rest frame limit $|{\bf P}| \to 0$, so it is excluded.
The second one brings the spin operator in the form
\begin{equation}
S^a = P_5^{-1} {W_5}^a.
\label{spinoperator}
\end{equation}
In terms of the Galilei group generators, the spin operator is written as
\begin{equation}
{\bf S} = {\bf J} - P_5^{-1} [{\bf K} \times {\bf P}].
\label{vectorspin}
\end{equation}

From the relations (\ref{relation2})-(\ref{relation3}), we obtain
\begin{eqnarray}
{[S^a, P^{\mu}]}_{-} & = & 0, \nonumber \\
{[J^a, S^b]}_{-} & = & {\rm i}{\epsilon}^{abc}S^c, \nonumber
\end{eqnarray}
i.e. $S^a$ behaves like a $3$-dimensional vector under rotations, being invariant under
translations.

In contrast with the spin operator in Lorentz covariant theories, which transforms in a
noncovariant way under Lorentz boosts, the spin operator given by Eq.(\ref{vectorspin})
is invariant under Galilean boosts as well,
\[
{[K^a, S^b]}_{-} = 0.
\]

An alternative way of introducing the Lorentz covariant spin operator is by performing
a transformation to the rest frame and identifying the spin components as spatial part
of the axial vector $W_{5{\mu}}$. Along the same lines, we perform the
transformation
\begin{eqnarray}
W_{5\mu}^{\prime} & = & {G_{\mu}}^{\lambda} W_{5\lambda},
\label{wtransform} \\
P_{\mu}^{\prime} & = & {G_{\mu}}^{\lambda} P_{\lambda},
\label{ptransform}
\end{eqnarray}
where ${G_{\mu}}^{\lambda}$ is a Galilean boost matrix chosen in such a way that
\[
P_a^{\prime} =0.
\]
For ${G_{\mu}}^{\lambda}$, we obtain
\begin{eqnarray}
{G_a}^b = {{\delta}_a}^b, & \qquad & {G_a}^5= - {G_4}^a = - \dfrac{P_a}{P_5}, \nonumber \\
{G_4}^5 = \dfrac{{\bf P}^2}{2P_5^2}, & \qquad & {G_4}^4 = {G_5}^5 = 1,
\label{boost} \\
{G_5}^4 = 0, & \qquad & {G_a}^4 = {G_5}^a = 0. \nonumber
\end{eqnarray}
Using (\ref{boost}) in equation (\ref{ptransform}), this yields
\[
P_4^{\prime} = P_4^{(0)},
\]
demonstrating that (\ref{boost}) is a transformation to the rest frame.

From equation (\ref{wtransform}), we also have
\begin{eqnarray}
W_{54}^{\prime} & = & 0, \nonumber \\
W_{5a}^{\prime} & = & W_{5a}, \nonumber
\end{eqnarray}
i.e $W_{5a}$ is invariant under Galilean boosts. The operator
\[
W_{5{\mu}}^{\prime} = ( W_{5a}, 0, 0),
\]
has only three nonvanishing components, and we use them to introduce
the spin operator given again by Eq.(\ref{spinoperator}).

The invariant $I_3$ becomes
\[
I_3 = P_5^{-2} S_a S^a,
\]
so that $P_5^2 I_3$ is related to the magnitude of spin.

Multiplying both sides of Eq.(\ref{vectorspin}) by ${\bf P}$, we get the relation ${\bf P}{\bf S} =
{\bf P}{\bf J}$ which holds in the zero inertial mass subspace as well. It shows that the component
of the angular momentum ${\bf J}$ along the direction of the linear momentum ${\bf P}$, i.e helicity,
can be defined in all subspaces.
\subsection{Galilean Dirac Lagrangian}

The five-dimensional Dirac field model is given by the Lagrangian density:
\begin{equation}
 {\cal L}_0(x)= \overline{\Psi}(x)({\rm i} {\gamma}^{\mu}
{{\partial}_{\mu}}-k)\Psi(x), \label{lagdirac}
\end{equation}
where
 $k$ is given by Eq.(\ref{defk}), and ${\Psi}(x)$ is a free Dirac field defined on the five-dimensional manifold ${{\cal G}}_{\rm (4+1)}$ with the Galilean metric in Eq.(\ref{galileanmetric}).
The Dirac matrices $\gamma^{\mu}$ in the extended space-time are four-dimensional and obey the usual anti-commutation relations:
\[
\left\{ \gamma^{\mu},\gamma^{\nu}\right\}=2g^{\mu \nu}.
\]
We use the following representation, which is convenient for a particle-antiparticle interpretation \cite{marc}:
\[
\gamma^a=
\left(\begin{array}{cc}
0 & {\rm i}\sigma^a\\
{\rm i}\sigma^a & 0
\end{array}\right),\;\;\;
\gamma^{4}=\frac{1}{\sqrt{2}}\left(
\begin{array}{cc}
1&1\\
-1&-1
\end{array}\right),
\]
\[
\gamma^5=\frac{1}{\sqrt{2}} \left(
\begin{array}{cc}
1&-1\\
1&-1
\end{array}\right),
\]
where ${\sigma}^1$,${\sigma}^2$, and ${\sigma}^3$ are the usual Pauli matrices.
The adjoint field is defined as
\[
\overline{\Psi}(x)=\Psi^\dagger(x)\ \gamma^0,\]
where
\[
\gamma^0=\frac{1}{\sqrt{2}}\left(\gamma^4+\gamma^5\right)=
\left(
\begin{array}{cc}
1 & 0\\
0 & -1
\end{array}
\right).
\]

The Euler-Lagrange equations of motion for $\Psi(x)$ and its adjoint $\overline{\Psi}(x)$, respectively, are
\begin{equation}
 \left({\rm i}
\gamma^{\mu}\partial_{\mu}-k\right)\Psi(x)=0,\qquad
\overline{\Psi}(x) ({\rm i}
\gamma^\mu\stackrel{\leftarrow}{\partial}_\mu+k)=0,\label{d1}
\end{equation}
where ${\zeta} \stackrel\leftarrow\partial {\chi}=(\partial {\zeta}){\chi}$.
The action corresponding to the Lagrangian density in Eq.(\ref{lagdirac})
is assumed to be invariant with respect to the inhomogeneous Galilean transformations of coordinates.

A transition from one reference frame $x$ to another $x'$ through a Galilean transformation
corresponds to a homogeneous linear transformation of the Dirac field,
\begin{equation}
{\Psi}(x) \to {\Psi'}(x')={\Lambda} {\Psi}(x),
\label{psitrans}
\end{equation}
where ${\Lambda}$ is determined by the parameters of the Galilean transformation.

In the manner of Lorentz covariant theories, ${\Lambda}=1$ for space-time  translations and for
 translations in the $x^5$-direction. For spatial rotations, we have
\begin{equation}
{\Lambda}=\exp\left\{ \frac{1}{2} {\gamma}^a {\gamma}^b {\epsilon}^{abc} {\phi}^c \right\},
\label{rotations}
\end{equation}
where $({\phi}^1, {\phi}^2, {\phi}^3)$ are angles of rotation.

Now we compute ${\Lambda}$ for Galilean boosts; that is, $R=I$ in Eq.(\ref{strans}). From Eq.(\ref{strans}), we find the Galilean boost transformations for the derivatives:
\begin{eqnarray}
{\partial}^{\prime}_{a} & = & {\partial}_a - {\beta}_a {\partial}_5, \nonumber \\
{\partial}^{\prime}_{4} & = & {\partial}_4 - {\beta}_a {\partial}_{a} + \frac{1}{2} |{\boldsymbol{\beta}}|^2 {\partial}_5, \\
{\partial}^{\prime}_{5} & = & {\partial}_5, \nonumber
\label{dertrans}
\end{eqnarray}
where $( {\partial}_a, {\partial}_4, {\partial}_5 )$ and $( {\partial}^{\prime}_a, {\partial}^{\prime}_4, {\partial}^{\prime}_5 )$ are the derivatives with respect to the original and
 the new coordinates, respectively.

According to Galilean covariance, the Dirac equation must retain its form in terms of the transformed coordinates and fields; that is,
\begin{equation}
( {\rm i} {\gamma}^{\mu} {\partial}^{\prime}_{\mu} - k) {\Psi}^{\prime}(x')=0.
\label{diraceq}
\end{equation}
By multiplying  Eq. (\ref{diraceq}) on the left by ${\Lambda}^{-1}$, and
 by  using Eqs. (\ref{psitrans}) and (\ref{dertrans}), we obtain the following relations:
\begin{eqnarray}
{\Lambda}^{-1} {\gamma}^a {\Lambda} & = & {\gamma}^a - {\beta}^a {\gamma}^4, \nonumber \\
{\Lambda}^{-1} {\gamma}^4 {\Lambda} & = & {\gamma}^4, \nonumber \\
{\Lambda}^{-1} {\gamma}^5 {\Lambda} & = & {\gamma}^5 + {\beta}^a {\gamma}^a + \frac{1}{2} |{\boldsymbol{\beta}}|^2 {\gamma}^4. \nonumber
\end{eqnarray}
From these relations we find the transformation matrix for Galilean boosts in the form
\begin{equation}
{\Lambda}=\exp\left\{ \frac{1}{2} {\gamma}^4 {\gamma}^a {\beta}_a \right\} =
 1 + \frac{1}{2} {\gamma}^4 {\gamma}^a {\beta}_a,
\label{boost1}
\end{equation}
where we have used $({\gamma}^4)^2=0$.

\subsection{Non-Galilean boosts}

The Lagrangian density given by Eq.(\ref{lagdirac}) preserves its form under the replacement
\begin{equation}
x^4 \leftrightarrow x^5, \qquad {\gamma}^4 \leftrightarrow {\gamma}^5.
\label{replacement}
\end{equation}
This symmetry reflects itself in the invariance of the action with respect to non-Galilean transformations of the form
\begin{eqnarray}
{\bf x'} &=& {\bf x} - {\boldsymbol{\tau}} x^5,\nonumber\\
{x'}^{4}&=&x^4-({\bf x} \cdot {\boldsymbol{\tau}}) +\frac{1}{2}{|{\boldsymbol{\tau}}|}^{2} x^5,
\label{x5trans} \\
{x'}^{5}&=&x^5, \nonumber
\end{eqnarray}
the transformation matrix for the Dirac field becoming
\begin{equation}
{\Lambda}= 1 + \frac{1}{2} {\gamma}^5 {\gamma}^a {\tau}_a.
\label{x5boost}
\end{equation}
In accordance with (\ref{replacement}), the transformation (\ref{x5trans}) can be obtained from the Galilean one by applying $x^4 \leftrightarrow x^5$, and the transformation matrix given by Eq.(\ref{x5boost}) results from Eq.(\ref{boost1}) by replacing ${\gamma}^4$ with ${\gamma}^5$. The parameters ${\tau}_a$ play the same role as the parameters ${\beta}_a$ in the Galilean boosts,
\[
{\tau}^a = \frac{dx^a}{dx^5} = \bar{c} \frac{dx^a}{ds},
\]
where the "velocity" $dx^a/ds$ determines the rate of change of $x^a$ in the fifth coordinate $s$.

Although $I_1$ is invariant with respect to the transformation (\ref{x5trans}) as well, the invariance of $I_2$ and $I_3$ is violated. In particular, the energy of the system in two reference frames connected by (\ref{x5trans}) is the same, $E^{\prime} = E$, while the inertial mass transforms as follows
\begin{equation}
m \to m^{\prime}= m - \frac{({\boldsymbol{\tau}} {\bf p})}{\bar{c}} +
\frac{1}{2}{|{\boldsymbol{\tau}}|}^{2} \frac{E}{{\bar{c}}^2}.
\label{masstrans}
\end{equation}
Assuming that $\bar{c}$ is the same in all reference frames, either Galilean or non-Galilean, we conclude from Eq.(\ref{constc}) that
\begin{equation}
\frac{m^{\prime}}{m} = \frac{m_0}{m^{\prime}_0},
\label{mrelation}
\end{equation}
i.e. the inertial and rest masses are transformed in such way that the product $m m_0$ remains unchanged. This gives us the following expression for the transformed rest mass $m^{\prime}_0$ in terms of the original $m$ and $m_0$:
\begin{equation}
m^{\prime}_0 = m_0 \Big( 1 - \frac{({\boldsymbol{\tau}} {\bf p})}{k} \sqrt{\frac{2m_0}{m}} +
\frac{1}{2}{|{\boldsymbol{\tau}}|}^{2} \Big(1 + \frac{{\bf p}^2}{k^2}\Big) \frac{m_0}{m} \Big)^{-1}.
\label{restmasstrans}
\end{equation}
The transformation (\ref{x5trans}) can be referred to as a boost in the ${\bf x}$-direction with velocity $d{\bf x}/ds$ or a non-Galilean boost. A general form of non-Galilean transformations includes rotations in the way similar to Eq.(\ref{strans}).

\subsection{Conserved currents and generators}

As a consequence of the invariance of the action with respect to translations, our model has five conserved currents
grouped together as an {\em energy-momentum-mass} tensor $T^{{\mu}{\nu}}$:
\begin{equation}
{\partial}_{\mu} T^{{\mu}{\nu}}=0,
\label{currents}
\end{equation}
where
\[
T^{{\mu}{\nu}}= \overline{\Psi}(x){\rm i} \gamma^\mu
{\partial^\nu}\Psi(x) - {\cal L}_0 g^{{\mu}{\nu}}.
\]
By integrating Eq.(\ref{currents}) over a volume limited in $x^4$-direction by two four-dimensional
hyper-surfaces, $x^4={\rm constant}$, and of infinite extent in other directions, and by assuming that the fields vanish at the infinite boundaries, we obtain the time-independent five-momentum of the Galilean Dirac field:
\[
P_{\mu} = \int d^3x dx^5 \sqrt{2} \; T_{5{\mu}}.
\]
This gives the explicit expressions for generators $P_4$ and $P_5$:
\begin{eqnarray}
P_4 =  \int d^3x  dx^5  \sqrt{2} \; \big( & - \overline{\Psi}(x)({\rm i} \gamma^a
{\partial_a}-k)\Psi(x) \nonumber \\
& - \overline{\Psi}(x) {\rm i} \gamma^5
{\partial_5}\Psi(x) \big)
\label{t54}
\end{eqnarray}
and
\begin{equation}
P_5 = \int d^3x  dx^5 \sqrt{2} \; \overline{\Psi}(x) {\rm i} \gamma^4
{\partial_5}\Psi(x).
\label{t55}
\end{equation}
While the first term in the right-hand side of Eq.(\ref{t54}) represents
the kinetic and mass term contributions to the total energy, which have the same
form as in $4$-dimensional relativistic theories, the $x^5$ dependence introduces
a new type of interaction and therefore a new type of energy contribution.

We rescale the generators $P_4$ and $P_5$ and define the Hamiltonian $H$ and the total
mass $M$ of the system as follows
\begin{equation}
\bar{c} P_4 = H, \qquad  \dfrac{1}{\bar{c}} P_5 = M,
\label{rescaling}
\end{equation}

As in the case of Lorentz covariant theories, we can introduce the Belinfante tensor
\begin{equation}
{\Theta}^{\mu \nu} = T^{\mu \nu} - \frac{1}{2} {\partial}_{\rho}
[ \overline{\Psi}(x) F^{\rho \mu \nu} \Psi(x)],
\end{equation}
where
\[
F^{\rho \mu \nu} \equiv {\gamma}^{\rho} {\cal F}^{\mu \nu}
- {\gamma}^{\mu} {\cal F}^{\rho \nu} - {\gamma}^{\nu} {\cal F}^{\rho \mu}
\]
and
\[
{\cal F}^{\mu \nu} = - \frac{\rm i}{4} [{\gamma}^{\mu}, {\gamma}^{\nu}]_{-},
\]
which satisfies the same conservation law as $T^{\mu \nu}$,
\begin{equation}
{\partial}_{\mu} {\Theta}^{\mu \nu} = 0,
\label{conserv}
\end{equation}
and provides us with the same expression for the time-independent five-momentum:
\begin{equation}
P^{\Theta}_{\mu} \equiv \int d^3x dx^5 \sqrt{2} {\Theta}_{5 \mu} = P_{\mu}.
\label{newmomentum}
\end{equation}
This means that ${\Theta}^{\mu \nu}$ can also be regarded as the energy-momentum-mass tensor. The explicit expression for ${\Theta}^{\mu \nu}$ is
\begin{equation}
{\Theta}^{\mu \nu}= \frac{1}{2} \overline{\Psi}(x) i ({\gamma}^{\mu} \stackrel{\leftrightarrow}{{\partial}^{\nu}}
+ {\gamma}^{\nu} \stackrel{\leftrightarrow}{{\partial}^{\mu}}) {\Psi}(x)- {\cal L}_0 g^{\mu \nu},
\label{teta}
\end{equation}
where $a\stackrel{\leftrightarrow}{\partial b}\equiv \frac 12\left [a\partial b - (\partial a)b\right ]$,  ${\Theta}^{\mu \nu}$ being symmetric, ${\Theta}^{\mu \nu} = {\Theta}^{\nu \mu}$.

The invariance of the action with respect to spatial rotations and boosts, gives us another set of conserved currents:
\begin{equation}
{\partial}_{\mu} M^{{\mu},{\rho}{\sigma}}=0,
\label{conserved}
\end{equation}
where
\begin{equation}
M^{{\mu},{\rho}{\sigma}} \equiv x^{\rho} {\Theta}^{\mu \sigma} - x^{\sigma} {\Theta}^{\mu \rho}.
\label{tensor}
\end{equation}
The conservation of $M^{{\mu},{\rho}{\sigma}}$ is provided by the symmetry of ${\Theta}^{\mu \nu}$. Integrating Eq.(\ref{tensor}) in the same way as in the case of translations, we obtain the time-independent tensor
\begin{equation}
M^{\rho \sigma} = \int d^3x dx^5 \sqrt{2} \; M^{4,\rho \sigma}.
\label{timeind}
\end{equation}

For spatial components of indices ${\rho}$ and ${\sigma}$, $M^{ab}$ is the angular momentum tensor related to rotations in the $x^ax^b$-planes. As in Lorentz covariant models, $M^{ab}$ consists of two parts
\[
M^{ab} = M_{0}^{ab} + {\cal S}^{ab},
\]
where
\[
M_{0}^{ab} \equiv \int d^3x dx^5 \sqrt{2} \; (x^a T^{4b} - x^b T^{4a})
\]
is the orbital part of $M^{ab}$ , while
\[
{\cal S}^{ab} \equiv \int d^3x dx^5 \; \sqrt{2} \bar{\Psi}(x) {\gamma}^{4} {\cal F}^{ba} {\Psi}(x)
\]
is its spin part.

The tensor $M^{ab}$ has three independent components, that can be represented by the vector $J^a$ in (\ref{commutation}),
\[
J^a \equiv J_0^a + {\cal S}^a,
\]
where $J_0^a =(1/2) {\epsilon}^{abc} M_0^{bc}$. The spin part of this vector is
\[
{\bf \cal S} = \frac{1}{\sqrt{2}} \int d^3x  dx^5\; \bar{\Psi}(x) {\gamma}^4 {\bf \Sigma} {\Psi}(x),
\]
which is often called the Dirac spin operator,
where
\[
{\Sigma}^a =
\left(\begin{array}{cc}
\sigma^a & 0\\
0 & \sigma^a
\end{array}\right) =
\dfrac{\rm i}{2} {\gamma}^a ({\gamma}^4 {\gamma}^5 - {\gamma}^5 {\gamma}^4).
\]
The Galilean covariant and Dirac spin operators are related as
\begin{equation}
{\bf S} = {\bf \cal S} - P_5^{-1} [{\bf K} \times {\bf P}] + {\bf J}_0.
\label{relation}
\end{equation}
Like ${\bf S}$, the Dirac spin operator ${\bf \cal S}$ is invariant under Galilean boosts. In particular, under
the transformation (\ref{boost}) ${\Sigma}^a$ transforms as follows
\[
{\Sigma}^a \to {\Lambda}^{-1} {\Sigma}^a {\Lambda} =
{\Sigma}^a + {\rm i}{\gamma}^4 {\epsilon}^{abc}{\gamma}^b \frac{P^c}{2P_5},
\]
where ${\Lambda}$ is given by Eq.(\ref{boost1}) with $v_a=P_a/P_5$, while ${\bf \cal S}$ does not change. In the
rest frame, the last two terms in the right-hand side of Eq.(\ref{relation}) vanish, so the operators ${\bf S}$
and ${\bf \cal S}$ coincide. However, there is an important difference between the two spin operators. The Dirac spin is not in general conserved in time. This is specific for Galilean covariant theories in $(4+1)$ dimensions.

When we set ${\rho}=a$ and ${\sigma}=4$ or vice versa, Eq.(\ref{timeind}) gives us the Galilean boost generator
\begin{equation}
K^a = M^{a4} = x^4 P_a - \int d^3x dx^5\; \sqrt{2} x_a T_{55}.
\label{galboost}
\end{equation}
The spin part of $M^{a4}$ is
\[
S^{a4} = \frac{1}{\sqrt{2}} \int d^3x dx^5\; \bar{\Psi}(x) F^{a44} {\Psi}(x).
\]
Since $F^{a44}=0$, it vanishes and does not contribute to Eq.(\ref{galboost}).

\subsection{Nonconservation of Dirac spin}

The remaining nonzero components of the tensor (\ref{timeind}) are $M^{45}$ and $M^{a5}$. The conservation of
\begin{equation}
M^{45} = x^4 P_4 - \int d^3x dx^5 \sqrt{2} \; x^5 {\Theta}_{55}
\label{m45}
\end{equation}
is another form of Eq.(\ref{newmomentum}) for ${\mu}=4$. Using Eq.(\ref{conserv}), we can rewrite $P_4$ as
\[
P_4 = \frac{d}{dx^4} \int d^3x dx^5 \sqrt{2} \; x^5 {\Theta}_{55}.
\]
Integrating this expression over $x^4$, this gives us the time-independent quantity represented by Eq.(\ref{m45}).

The component $\tilde{K}^a \equiv M^{a5}$ is a generator of non-Galilean boosts. In contrast with the Galilean boost generator, it has a non-vanishing spin part
\begin{equation}
\tilde{\cal S}^a \equiv {\cal S}^{a5} = \dfrac{1}{\sqrt{2}} \int d^3x dx^5 \bar{\Psi}(x) {\gamma}^4 {\tilde{\Sigma}}^a {\Psi}(x),
\label{newspin}
\end{equation}
where $\tilde{\Sigma}^a \equiv - {\Sigma}^a {\gamma}^5$.

The nonzero commutation relations including $\tilde{K}^a$ are
\begin{eqnarray}
{[J^a, \tilde{K}^b]}_{-} & = & {\rm i} {\epsilon}^{abc} \tilde{K}^c, \qquad  {[P^a, \tilde{K}^b]}_{-}= {\rm i} g^{ab} P_4, \nonumber\\
{[\tilde{K}^a, P_5]}_{-} & = & {\rm i} P^a \nonumber
\end{eqnarray}
(compare to the corresponding commutators in (\ref{commutation})). In addition, we get
\[
{[K^a, \tilde{K}^b]}_{-} = -{\rm i} {\epsilon}^{abc} J^c - {\rm i} g^{ab} \tilde{Q},
\]
i.e. the galilean and non-Galilean boosts generators do not commute, $\tilde{Q} \equiv M^{45}$ playing the role of central charge. The nonzero commutation relations for $\tilde{Q}$ are
\begin{eqnarray}
{[\tilde{Q}, K^a]}_{-} = & {\rm i} K^a, \qquad   {[\tilde{Q}, \tilde{K}^a]}_{-} & =   -{\rm i} \tilde{K}^a \nonumber\\
{[\tilde{Q}, P_5]}_{-} = & {\rm i} P_5 \qquad   {[\tilde{Q}, P_4]}_{-}& = -{\rm i} P_4. \nonumber
\end{eqnarray}
The nonconservation in time of the Dirac spin is represented as
\begin{equation}
\frac{d}{dx^4} {\bf \cal S}= \dfrac{1}{\sqrt{2}} \int d^3x dx^5 \bar{\Psi} \Big( {\gamma}^4 [{\bf \tilde{\Sigma}} \times {\bf \nabla}] - [{\bf \tilde{\Sigma}} \times {\bf \nabla}] {\gamma}^4 \Big) {\Psi},
\label{noncons}
\end{equation}
being related to the spin part of the non-Galilean boost generator.

The $x^4 \leftrightarrow x^5$ symmetry and the role of $x^5$ coordinate become more obvious if we represent the four-component spinor ${\Psi}(x)$ as a set of two two-component spinors ${\Psi}_1(x)$ and ${\Psi}_2(x)$ as follows
\[
{\Psi}(x)=
\left(
\begin{array}{c}
{\Psi}_{1}(x) \\ {\Psi}_{2}(x)
\end{array}
\right).
\]
For the linear combinations of these spinors
\begin{eqnarray}
{\Xi}(x) & = &
\left(
\begin{array}{c}
{\Xi}_{1}(x) \\ {\Xi}_{2}(x)
\end{array}
\right)\nonumber \\
& \equiv &
\left(
\begin{array}{c}
{\Psi}_{1}(x) + {\Psi}_{2}(x) \\ {\Psi}_{1}(x) - {\Psi}_{2}(x)
\end{array}
\right),
\label{lincom}
\end{eqnarray}
the Dirac equation given in Eq.(\ref{d1}) reduces to the system of two equations
\begin{eqnarray}
{\partial}_4 {\Xi}_{1} & = &  {\rm i} p_{-} {\Xi}_{2}, \nonumber \\
{\partial}_5 {\Xi}_{2} & = & - {\rm i} p_{+} {\Xi}_{1},
\label{newsys}
\end{eqnarray}
where
\[
p_{\pm} \equiv \dfrac{1}{\sqrt{2}} ({\sigma}^a {\partial}_a \pm k).
\]
We see that the second component ${\Xi}_2$ is not dynamically independent. Its time evolution is completely determined by ${\Xi}_1$, while the $x^5$-dependence of the first component ${\Xi}_1$ is governed by ${\Xi}_2$. Acting at both sides of these equations by ${\partial}_5$ and ${\partial}_4$, respectively, this gives us
\begin{eqnarray}
{\partial}_4 {\partial}_5 {\Xi}_1 & = & p_{-}p_{+} {\Xi}_1, \nonumber \\
{\partial}_4 {\partial}_5 {\Xi}_2 & = & p_{-}p_{+} {\Xi}_2, \nonumber
\end{eqnarray}
i.e. the second order form of these equations is the same.

The generators of the Galilei Lie algebra can be expressed in terms of ${\Xi}_1$ only. For instance, the generators of translations and the spin operator take the form
\[
P_{\mu} = \int d^3x dx^5 {\Xi}_1^{\dagger} {\rm i} {\partial}_{\mu} {\Xi}_1
\]
and
\[
{\cal S}^a = \dfrac{1}{2} \int d^3x dx^5  {\Xi}_1^{\dagger} {\sigma}^a {\Xi}_1.
\]
However, ${\Xi}_2$ contributes to the spin part of the non-Galilean boost generator:
\begin{eqnarray}
\tilde{\cal S}^a & = & - \dfrac{1}{\sqrt{2}} \int d^3x dx^5 {\Xi}_1^{\dagger} {\sigma}^a {\Xi}_2 \nonumber \\
& = & \dfrac{1}{\sqrt{2}} \int d^3x dx^5 {\Xi}_1^{\dagger} {\sigma}^a p_{+} {\rm i} {\partial}_5^{-1}
{\Xi}_1 \nonumber
\end{eqnarray}
making its density nonlocal in $x^5$.

Equation (\ref{noncons}) becomes
\[
\frac{d}{dx^4} {\cal S}^a = \dfrac{1}{\sqrt{2}} \int d^3x dx^5 {\partial}_5 \Big( {\partial}_5^{-1} {\Xi}_1^{\dagger} {\sigma}^a p_{-}p_{+} {\partial}_5^{-1} {\Xi}_1 \Big).
\]
Due to the nonlocality in $x^5$, $\frac{d}{dx^4} {\cal S}^a$ does not vanish  unless the $x^5$ coordinate is compactified and/or boundary conditions on ${\partial}_5^{-1} {\Xi}_1$ are imposed.

\section{Reduction to $(3+1)$ dimensions}

The reduction to $(3+1)$ dimensions is performed by factoring the $x^5$ coordinate out of the original field
 ${\Psi}(x)$ as \cite{khan}
\begin{equation}
\Psi (x) = e^{-{\rm i} m \bar{c} x^5}{\psi}_{+}({\bf x},t).
\label{factor}
\end{equation}
The field ${\psi}_{+}({\bf x},t)$ is a nonrelativistic Fermi field  which represents particles with positive inertial mass $m$ in $(3+1)$ dimensions. The integral over $x^5$, for instance in Eqs.(\ref{t54}) and (\ref{t55}), is then interpreted as
 $\int dx^5 \to {\lim}_{l \to \infty} (1/l) \int_{-l/2}^{l/2} dx^5$, where $l$ is an arbitrary length, i.e. the $x^5$ coordinate is first compactified and factorized and then integrated out.

In two reference frames connected by the reduced Galilean transformation $({\bf x},x^4) \to ({\bf x}^{\prime},{x^{\prime}}^4)$, the fields ${\psi}_{+}^{\prime}({\bf x}^{\prime},t^{\prime})$ and ${\psi}_{+}({\bf x},t)$ are related as
\begin{equation}
{\psi}_{+}^{\prime}({\bf x}^{\prime},t^{\prime}) = e^{-{\rm i}{\Delta}({\bf x},t)} {\Lambda}
{\psi}_{+}({\bf x},t),
\label{phaserelation}
\end{equation}
where, in addition to the transformation matrix $\Lambda$ discussed above, we have a space-time-dependent phase factor with
\begin{equation}
{\Delta}({\bf x},t) \equiv  m\bar{c} \Big[ ({\bf Rx}) \cdot {\boldsymbol{\beta}} - \frac{1}{2}{|{\boldsymbol{\beta}}|}^{2} x^4 \Big].
\label{phasefactor}
\end{equation}
This phase factor is caused by the nontrivial cohomology of the Galilei group \cite{levy},\cite{azca}.

The ansatz (\ref{factor}) breaks the invariance of the original theory with respect to non-Galilean boosts. Assuming for a moment that the field  ${\Psi}(x)$ in (\ref{factor}) transforms under non-Galilean boosts in accordance with that invariance, i.e. like in Eq.(\ref{psitrans}) with $\Lambda$ given by Eq.(\ref{x5boost}), this yields
\[
{\psi}_{+}^{\prime}({\bf x}^{\prime},t^{\prime}) =
e^{{\rm i}(m^{\prime}-m)\bar{c} x^5} {\Lambda} {\psi}_{+}({\bf x},t)
\]
or
\[
{\partial}_5  {\psi}_{+}^{\prime}({\bf x}^{\prime},t^{\prime}) =
{\rm i} (m^{\prime}-m) \bar{c} {\psi}_{+}^{\prime}({\bf x}^{\prime},t^{\prime}),
\]
where $({\bf x}^{\prime},t^{\prime})$ and $({\bf x},t)$ are related by the transformation (\ref{x5trans}), while $m^{\prime} \neq m$ according to Eq.(\ref{masstrans}). Therefore, even if we start with a Galilean reference frame in which ${\psi}_{+}({\bf x},t)$ is $x^5$ independent, non-Galilean boosts bring us to a non-Galilean one where the factorization (\ref{factor}) does not hold.

The ansatz (\ref{factor}) is not invariant under translations in the $x^5$-direction as well. However, the factor
$\exp\{-{\rm i} m\bar{c}d\}$, which appears in Eq.(\ref{factor}) after the translation
   $x^5 \to x^5 + d$, can be removed by performing global phase transformation:
\[
{\psi}_{+}({\bf x},t) \to e^{{\rm i}m\bar{c}d} {\psi}_{+}({\bf x},t).
\]
With the factorization given by Eq.(\ref{factor}), the total mass of the system is
\[
M = m N_{+}.
\]
The symbol $N_{+}$,
\begin{eqnarray}
N_{+} & \equiv & \int d^3x {\cal N}_{+}({\bf x},t) \nonumber \\
 & = & \int d^3x \sqrt{2} \bar{\psi}_{+}({\bf x},t) {\gamma}^4 {\psi}_{+}({\bf x},t)
\label{numpart}
\end{eqnarray}
is the number of particles. We assume that the particles have the electric charge $+1$, so that $N_{+}$ stands for both the total number of particles and the total electric charge.

By substituting this factorization into the action, we rewrite the Lagrangian density of Eq.(\ref{lagdirac}) as
\begin{eqnarray}
 {\cal L}_0(x) & \to & {\cal L}_{0,+}({\bf x},t)\nonumber \\
& = & \overline{\psi}_{+}({\bf x},t)( {\rm i} \gamma^{\bar{\mu}}
\stackrel{\leftrightarrow}{\partial_{\bar{\mu}}}- kI_{+})\psi_{+}({\bf x},t), \label{newlagdirac}
\end{eqnarray}
where $\bar{\mu}$ runs from $1$ to $4$, and the matrix $I_{+}$ is defined by
\[
I_{\pm}=I \mp \frac{m\bar{c}}{k} {\gamma}^5,
\]
$I$ being the identity matrix. Below we will also use the matrix $I_{-}$ that differs from $I_{+}$ in the sign of m. Eq.(\ref{newlagdirac}) is the effective Lagrangian density in $(3+1)$ dimensions. The corresponding action is invariant under global phase transformations of the field ${\psi}_{+}({\bf x},t)$ and under reduced Galilean transformations of coordinates $({\bf x},x^4) \to ({\bf x}^{\prime},{x^{\prime}}^4)$.

\subsection{Non-relativistic Fermi field in $(3+1)$ dimensions}

The Euler-Lagrange equation of motion for the field ${\psi}_{+}({\bf x},t)$ is
\begin{equation}
\left({\rm i}
\gamma^{\bar{\mu}}\partial_{\bar{\mu}}-kI_{+}\right){\psi}_{+}({\bf x},t)=0.
\label{nonreleq}
\end{equation}
Representing the field ${\psi}_{+}({\bf x},t)$ as
\[
{\psi}_{+}({\bf x},t)=
\left(
\begin{array}{c}
{\psi}_{1,+}({\bf x},t) \\ {\psi}_{2,+}({\bf x},t)
\end{array}
\right),
\]
and introducing the independent and dependent fields, ${\eta}_{1,+}({\bf x},t)$ and ${\eta}_{2,+}({\bf x},t)$, respectively, in the same way as above for the Dirac field in $(4+1)$ dimensions,
\begin{eqnarray}
{\eta}_{+}({\bf x},t) & = &
\left(
\begin{array}{c}
{\eta}_{1,+}({\bf x},t) \\ {\eta}_{2,+}({\bf x},t)
\end{array}
\right)\nonumber \\
& \equiv &
\left(
\begin{array}{c}
{\psi}_{1,+}({\bf x},t) + {\psi}_{2,+}({\bf x},t) \\ {\psi}_{1,+}({\bf x},t) - {\psi}_{2,+}({\bf x},t)
\end{array}
\right),
\label{linearcom}
\end{eqnarray}
we bring the system given by Eq.(\ref{nonreleq}) to the following form
\begin{eqnarray}
{\rm i} {\partial}_4 {\eta}_{1,+} +  p_{-} {\eta}_{2,+} & = & 0, \nonumber \\
p_{+} {\eta}_{1,+} - m\bar{c} {\eta}_{2,+} & = & 0,
\label{system1}
\end{eqnarray}
where
\[
p_{\pm} \equiv \dfrac{1}{\sqrt{2}} ({\sigma}^a {\partial}_a \pm k),
\]
i.e the independent field ${\eta}_{1,+}({\bf x},t)$ obeys a Schr\"{o}dinger type equation. For $k=0$, the rest mass is equal to zero, and this system coincides with the L\'evy-Leblond equations \cite{leblond}. In what follows, we will assume that $m=m_0$.

Another way of separating the independent and the dependent fields is by introducing the four-component spinors
\begin{equation}
{\phi}_{+} = \frac{1}{\sqrt{2}} I_{-} {\gamma}^4 {\psi}_{+}
\label{indep}
\end{equation}
and
\begin{equation}
{\chi}_{+} =  \frac{1}{2} I_{-} {\gamma}^4 {\gamma}^5 {\psi}_{+}
\label{dep}
\end{equation}
for the independent and dependent fields, respectively. The inverse transformation is
\begin{equation}
{\psi}_{+}  =  \frac{1}{4} {\gamma}^5 {\gamma}^4 I_{-} {\phi}_{+} +
\frac{1}{2\sqrt{2}} {\gamma}^4 I_{-} {\chi}_{+}.
\label{inverse}
\end{equation}
The equations of motion become
\begin{eqnarray}
{\rm i} {\partial}_4 {\phi}_{+} & = & - \frac{1}{2\sqrt{2}} \Big( {\rm i}{\gamma}^a {\Gamma} {\partial}_a
- k ({\gamma}^0 +I) \Big) {\chi}_{+}, \nonumber \\
\frac{k}{2} {\Gamma} {\chi}_{+} & = & \Big( {\rm i} {\gamma}^a {\partial}_a + \frac{k}{2} {\Gamma} \Big) {\phi}_{+},
\label{fourcomp}
\end{eqnarray}
where
\[
{\Gamma} \equiv \frac{1}{\sqrt{2}} ({\gamma}^4 - {\gamma}^5) + 2{\rm i}{\cal F}^{45},
\]
\[
{\Gamma}^2 = 0.
\]
The equations (\ref{fourcomp}) reproduce Eq.(\ref{system1}) if we take ${\phi}_{+}$ and ${\chi}_{+}$ in the form
\[
{\phi}_{+}=
\left(
\begin{array}{c}
{\eta}_{1,+} \\ 0
\end{array}
\right),
\qquad
{\chi}_{+}=
\left(
\begin{array}{c}
{\eta}_{2,+} \\ 0
\end{array}
\right).
\]
With the transformation, Eq.(\ref{linearcom}), and eliminating ${\eta}_{2,+}$ in favor of ${\eta}_{1,+}$, this gives us the Lagrangian density in the form
\[
{\cal L}_{0,+}({\bf x},t) = {\eta}_{1,+}^{\dagger}({\bf x},t) \Big( \dfrac{\rm i}{\sqrt{2}} {\partial}_4
+ \frac{p_{+}p_{-}}{k} \Big) {\eta}_{1,+}({\bf x},t)
\]
reflecting once more the Schr\"odinger nature of the independent field.

\subsection{Quantization}

As in Lorentz covariant models, the general solution to
 Eq.(\ref{nonreleq}) may be expanded in terms of the plane wave solutions as
  in Ref. \cite{marc}:
\[
{\psi}_{+}({\bf x},t) =  \frac{1}{(2\pi)^{3/2}} \sum_{r} \int d^3{p}\;
a^{(r)}({\bf p}) u^{(r)}(p) e^{-{\rm i}p_{\bar{\mu}}x^{\bar{\mu}}},
\]
where $r=1,2$, the scalar product $p_{\bar{\mu}}x^{\bar{\mu}}$ is
\[
p_{\bar{\mu}}x^{\bar{\mu}}=-{\bf p}{\bf x} + \frac{E}{\bar{c}} x^4,
\]
with  $E$ given by Eq.(\ref{dispersion}), and $a^{(r)}({\bf p})$ and $a^{{\dagger} (r)}({\bf p})$
are annihilation and creation operators of positive mass particles.

The positive-energy spinors $u^{(r)}(p)= u^{(r)}({\bf p}, E, m)$ obey the equations
\[
({\gamma}^{\bar{\mu}} p_{\bar{\mu}} - k I_{+}) u^{(r)}(p) =  0
\]
and the orthonormality conditions
\begin{equation}
\bar{u}^{(r)}(p) u^{(s)}(p)  =  {\delta}_{rs},
\label{orthonorm}
\end{equation}
which are solved by
\[
u^{(r)}(p) =   d_u ({\gamma}^{\bar{\mu}} p_{\bar{\mu}} + k I_{-})u^{(r)}(0),
\]
where
\[
d_u = \frac{1}{2k} {\left( \frac{4m{\bar{c}}^2}{E+3m{\bar{c}}^2} \right)}^{1/2}.
\]

The rest frame spinors $u^{(r)}(0)=u^{(r)}({\bf p}=0)$ are taken in the form
\begin{equation}
u^{(r)}(0)=
\left(
\begin{array}{c}
{\xi}^{(r)}(0)\\
0\\
\end{array}
\right),
\label{restframe}
\end{equation}
with
\[
{\xi}^{(1)}(0)=
\left(
\begin{array}{c}
1\\
0\\
\end{array}
\right), \qquad
{\xi}^{(2)}(0)=
\left(
\begin{array}{c}
0\\
1\\
\end{array}
\right).
\]

By using Eqs.(\ref{orthonorm})-(\ref{restframe}), we find
\[
\sum_{r} u^{(r)}(p) \bar{u}^{(r)}(p) =  \frac{1}{2k} ({\gamma}^{\bar{\mu}} p_{\bar{\mu}} + k I_{-}).
\]
Quantizing the field ${\psi}_{+}({\bf x},t)$ by assuming that the annihilation and creation operators
obey the anticommutation relations:
\[
\Big\{ a^{(r)}({\bf p}), a^{{\dagger} (s)}({\bf q}) \Big\} =
{\delta}_{rs} {\delta}({\bf p} - {\bf q}),
\]
we obtain at equal times $x^4=y^4=t$ the following anticommutator
\[
\Big\{ {\psi}_{+}({\bf x},t), \overline{\psi}_{+}({\bf y},t) \Big\} = D_{+}({\bf x}-{\bf y})
- \frac{1}{2k^2\sqrt{2}} {\gamma}^4 {\bf \nabla}^2{\delta}({\bf x}-{\bf y}),
\]
where
\[
D_{+}({\bf x}-{\bf y}) \equiv \frac{1}{2} ({\gamma}^0 + I) {\delta}({\bf x}-{\bf y})
+ \frac{1}{2k}{\rm i}{\gamma}^{a} {\partial}_{a} {\delta}({\bf x}-{\bf y}).
\]
The appearance of terms that do not vanish unless $k~\to~\infty$ is the contribution of the dependent field to the anticommutator. For the independent field we have
\[
\Big\{ {\eta}_{1,+}({\bf x},t), {\eta}_{1,+}^{\dagger}({\bf y},t) \Big\} =
{\delta}({\bf x}-{\bf y}),
\]
which is the standard equal-time anticommutation relation for Fermi fields.

\subsection{One-particle states}

The plane wave solution expansion for the independent field ${\eta}_{1,+}({\bf x},t)$ is
\[
{\eta}_{1,+}({\bf x},t) =  \frac{1}{(2\pi)^{3/2}} \sum_{r} \int d^3{p}\;
a^{(r)}({\bf p}) u^{(r)}_{+}(p) e^{-{\rm i}p_{\bar{\mu}}x^{\bar{\mu}}},
\]
where
\[
u^{(r)}_{+}(p) = u^{(r)}_{1}(p) + u^{(r)}_{2}(p) = d_u ({\rm i}{\sigma}^a p_a + 2k) {\xi}^{(r)}(0)
\]
and
\[
u^{(r)}_{+}(0) = {\xi}^{(r)}(0).
\]
The spinors  $u^{(r)}_{+}(p)$ obey the same orthonormality conditions as $u^{(r)}(p)$ in Eq.(\ref{orthonorm}).

Using the ansatz (\ref{factor}) in the expressions for generators constructed in the previous section in the same way as in the expression for the total mass, this gives us the corresponding generators in $(3+1)$ dimensions. In terms of the annihilation and creation operators, the generators $P^a$ and $H$, for instance, become
\[
P^a_{+} = \sum_{r} \int d^3{p}\; p^a a^{{\dagger} (r)}({\bf p}) a^{(r)}({\bf p})
\]
and
\[
H_{+} = \sum_{r} \int d^3{p}\; E_p a^{{\dagger} (r)}({\bf p}) a^{(r)}({\bf p})
\]
with
\[
E_p = \frac{{\bf p}^2}{2m}.
\]

The orbital angular momentum and Dirac spin are now separately conserved. For the corresponding operators, we get
\begin{eqnarray}
J_{0,+}^a = \sum_{r,s} \int d^3p\; a^{{\dagger} (r)}({\bf p})
\Big[ & - & {\rm i} {\delta}_{rs} {\epsilon}^{abc} p_b \frac{\partial}{{\partial} p_c}
\nonumber \\
& - & S_{2,+}^a(r,s;p) \Big] a^{(s)}({\bf p}) \nonumber
\end{eqnarray}
and
\[
{\cal S}_{+}^a = \sum_{r,s} \int d^3p\; a^{{\dagger} (r)}({\bf p})
\Big[ S_{1,+}^a(r,s) + S_{2,+}^a(r,s;p) \Big] a^{(s)}({\bf p}),
\]
respectively, where
\begin{equation}
S_{1,+}^a(r,s) \equiv \frac{1}{2} {\xi}^{{\dagger} (r)}(0) {\sigma}^a {\xi}^{(s)}(0),
\end{equation}
\begin{eqnarray}
S_{2,+}^a(r,s;p) & \equiv & d_u^2 {\xi}^{{\dagger} (r)}(0) \{ (p_ap_b{\sigma}^b
- {\bf p}^2 {\sigma}^a) \nonumber \\
& + &  k{\epsilon}^{abc} (p_c {\sigma}^b - p_b {\sigma}^c) \} {\xi}^{(s)}(0).
\end{eqnarray}
The numbers $S_{1,+}^a(r,s)$ can be represented as entries of the following $(2 \times 2)$-matrices in the $(r,s)$-space:
\[
S_{1,+}^1 =
\frac{1}{2} \left(\begin{array}{cc}
0 & 1\\
1 & 0
\end{array}\right),\;\;\;
S_{1,+}^2 =\frac{1}{2}\left(
\begin{array}{cc}
0 & -{\rm i}\\
{\rm i} & 0
\end{array}\right),
\]
\[
S_{1,+}^3 =\frac{1}{2} \left(
\begin{array}{cc}
1 & 0\\
0 & -1
\end{array}\right),
\]
while $S_{2,+}^a(r,s;p)$ satisfies
\[
p_a S_{2,+}^a(r,s;p) = 0.
\]
The $S_{2,+}^a(r,s;p)$ contributions to the orbital angular momentum and Dirac spin operators are opposite in sign, so that these contributions cancel each other in the expression for the total angular momentum.

For bilinear combinations of $S_{1,+}^a(r,s)$ and $S_{2,+}^a(r,s;p)$, we get the conditions
\begin{eqnarray}
\sum_{q} S_{1,+}^a(q,r) S_{1,+}^a(s,q) & = & \frac{3}{4} {\delta}_{rs},
\label{cond1}\\
\sum_{q} S_{2,+}^a(q,r;p) S_{2,+}^a(s,q;p) & = & 2 d_u^2 {\bf p}^2 {\delta}_{rs},
\label{cond2}
\end{eqnarray}
and
\begin{eqnarray}
\sum_{q} ( S_{1,+}^a(q,r) S_{2,+}^a(s,q;p) & + & S_{2,+}^a(q,r;p) S_{1,+}^a(s,q) )
\nonumber \\
& = &  - 2 d_u^2 {\bf p}^2 {\delta}_{rs}.
\label{cond3}
\end{eqnarray}
%

%
%
%
%

Let us now introduce one-particle states as
\begin{equation}
|r;{\bf p};+ \rangle \equiv a^{{\dagger} (r)}({\bf p}) |0;+ \rangle,
\end{equation}
where $|0;+ \rangle$ is the vacuum state that contains no particles,
\[
a^{(r)}({\bf p}) |0;+ \rangle = 0 \quad {\rm for} \quad {\rm all} \quad {\rm r}\quad {\rm and} \quad {\bf p},
\]
and belongs to the zero inertial mass subspace. Since
\[
M|r;{\bf p};+ \rangle = m|r;{\bf p};+ \rangle, \quad P_{+}^a|r;{\bf p};+ \rangle = p^a|r;{\bf p};+ \rangle,
\]
\[
H|r;{\bf p};+ \rangle = E_p|r;{\bf p};+ \rangle,
\]
the states $|r;{\bf p};+ \rangle$ are characterized by the mass $m$, the momentum ${\bf p}$ and the energy $E_p$.

On one-particle states, the Galilean covariant spin operator
\[
S_{+}^a = {\cal S}_{+}^a - \frac{1}{m\overline{c}} {\varepsilon}^{abc} K_{+}^b P_{+}^c + J_{0,+}^a,
\]
where $K_{+}^a$ is the Galilean boost generator in $(3+1)$ dimensions, reduces to the Dirac one,
\[
S_{+}^a |r;{\bf p};+ \rangle = {\cal S}_{+}^a |r;{\bf p};+ \rangle,
\]
so that we can use any of them. With the conditions given by Eqs.(\ref{cond1})-(\ref{cond3}), we get
\[
(S_{+}^a)^2 |r;{\bf p};+ \rangle = \frac{3}{4} |r;{\bf p};+ \rangle
\]
indicating that $|r;{\bf p};+ \rangle$ are spin-$1/2$ states.

For a given momentum ${\bf p}$, there are two different states $(r=1,2)$, corresponding to two possible values of the third component of spin. The action of $S_{+}^3$ on the states $|r;{\bf p};+ \rangle$ is as follows
\begin{eqnarray}
S_{+}^3 |1;{\bf p};+ \rangle & = & \Big(\frac{1}{2} - f_1\Big) |1;{\bf p};+ \rangle + f_2 |2;{\bf p};+ \rangle, \nonumber \\
S_{+}^3 |2;{\bf p};+ \rangle & = & -\Big(\frac{1}{2} - f_1\Big) |2;{\bf p};+ \rangle + f_2^{\star} |1;{\bf p};+ \rangle, \nonumber
\end{eqnarray}
where
\begin{eqnarray}
f_1 & = & \frac{p_1^2 + p_2^2}{{\bf p}^2 + 4k^2}, \nonumber \\
f_2 & = & \frac{(p_1 + {\rm i}p_2)(p_3 - {\rm i}2k)}{{\bf p}^2 + 4k^2} \nonumber
\end{eqnarray}
and
\[
f_2 f_2^{\star} = f_1(1 - f_1).
\]

For $p_1=p_2=0$, we have ${\bf p}=(0,0,p_3)$, $f_1=f_2=0$, and
\begin{eqnarray}
S_{+}^3 |1;p_3;+ \rangle & = & \frac{1}{2} |1;p_3;+ \rangle , \nonumber \\
S_{+}^3 |2;p_3;+ \rangle & = & -\frac{1}{2} |2;p_3;+ \rangle, \nonumber
\end{eqnarray}
so that the states $|1;p_3;+ \rangle$ and $|2;p_3;+ \rangle$ correspond to the values of $1/2$ and $(-1/2)$ of the third component of spin, respectively.

For $p_1 \neq 0$, $p_2 \neq 0$, $|1;{\bf p};+ \rangle$ and $|2;{\bf p};+ \rangle$ are not eigenstates of $S_{+}^3$. However, we can introduce their linear combinations
\begin{eqnarray}
|{\uparrow} ;{\bf p};+ \rangle & \equiv & \frac{1-f_1}{f_2} |1;{\bf p};+ \rangle + |2;{\bf p};+ \rangle, \nonumber \\
|{\downarrow};{\bf p};+ \rangle & \equiv & |1;{\bf p};+ \rangle - \frac{1-f_1}{f_2^{\star}} |2;{\bf p};+ \rangle \nonumber
\end{eqnarray}
which correspond to the same values of the third component of spin:
\begin{eqnarray}
S_{+}^3 |{\uparrow};{\bf p};+ \rangle & = & \frac{1}{2} |{\uparrow};{\bf p};+ \rangle , \nonumber \\
S_{+}^3 |{\downarrow};{\bf p};+ \rangle & = & -\frac{1}{2} |{\downarrow};{\bf p};+ \rangle. \nonumber
\end{eqnarray}
This is in agreement with the fact that for nonrelativistic particles the third component of spin can be used to enumerate the spin states for any value of momentum \cite{beres}.

According to Eqs.(\ref{phaserelation}) and (\ref{phasefactor}), in two Galilean reference frames connected by a pure Galilean boost the components of the momentum of a particle are related as follows
\[
{\bf p} \to {\bf p^{\prime}} = {\bf p} - m\bar{c} {\boldsymbol{\beta}}.
\]
Choosing $\boldsymbol{\beta}$ as
\[
\boldsymbol{\beta} = \Big( \frac{p_1}{m\bar{c}}, \frac{p_2}{m\bar{c}}, 0 \Big),
\]
we can always make a transition from a reference frame with $p_1 \neq 0$, $p_2 \neq 0$ to another one in which these two components of the momentum are zero and the enumeration of spin states is straightforward. This represents another way of identifying spin states for nonzero values of $p_1$ and $p_2$.

\subsection{Particle-antiparticle system}

We can study the system of $n$ particles of different mass as well. In this case, the theory includes $n$ sectors corresponding to masses $m_1,m_2,..,m_n$. The masses can be either positive (particles) or negative (antiparticles).

Let us consider a particle-antiparticle system with masses $m$ and $-m$, respectively. In addition to the field
${\psi}_{+}({\bf x},t)$ we get the field ${\psi}_{-}({\bf x},t)$ representing antiparticles. The Lagrangian density of the system becomes
\begin{eqnarray}
 {\cal L}_0({\bf x},t) & = & \sum_{i=(+,-)} {\cal L}_{0,i}({\bf x},t)\nonumber \\
& = & \sum_{i=(+,-)} \overline{\psi}_{i}({\bf x},t)( {\rm i} \gamma^{\bar{\mu}}
\stackrel{\leftrightarrow}{\partial_{\bar{\mu}}}- kI_{i})\psi_{i}({\bf x},t). \label{lagdirac2}
\end{eqnarray}
The mass superselection rule forbids the superposition of states of different mass in $(3+1)$ dimensions. However, we can use a linear combination of ${\psi}_{+}({\bf x},t)$ and ${\psi}_{-}({\bf x},t)$ in transition from $(4+1)$ dimensions to $(3+1)$ dimensions \cite{marc}.

As in the case with a single mass sector, we can start with the five-dimensional Dirac field model given by Eq.(\ref{lagdirac}) and then perform the reduction to $(3+1)$ dimensions by applying the factorization
\begin{equation}
\Psi (x) = e^{-{\rm i} m \bar{c} x^5}{\psi}_{+}({\bf x},t) + e^{{\rm i} m \bar{c} x^5}{\psi}_{-}({\bf x},t).
\label{factor2}
\end{equation}
By substituting this factorization into the corresponding action, we find that the particle and antiparticle fields contribute separately. The mixed terms contain the oscillating factors $e^{{\pm}2{\rm i}m\bar{c}x^5}$ so that their contribution to the action vanishes and the effective Lagrangian density in $(3+1)$ dimensions takes the form given in Eq.(\ref{lagdirac2}).

The conserved currents and generators, including the Dirac spin operator, contain two parts, one for positive mass sector and the other for negative one. For instance, ${\bf P} = {\bf P}_{+} + {\bf P}_{-}$, ${\bf K} = {\bf K}_{+} + {\bf K}_{-}$ and etc. The total mass of the system becomes
\[
M = m(N_{+}-N_{-}),
\]
where $N_{-}$ stands for the total number of antiparticles being given by Eq.(\ref{numpart}) with  ${\psi}_{+}({\bf x},t)$ replaced by ${\psi}_{-}({\bf x},t)$.

The exception is the Galilean covariant spin operator. The operators ${\bf S}_{+}$ and ${\bf S}_{-}$ are not additive \cite{gursey}, and in the case of nonzero total mass the total spin operator ${\bf S}$ is given by
\[
{\bf S} = \frac{P_{5,+}}{P_5} {\bf S}_{+} + \frac{P_{5,-}}{P_5} {\bf S}_{-} + {\bf J}_{0,rel},
\]
where
\[
P_5 = P_{5,+}+P_{5,-}, \qquad P_{5,\pm} = \pm m\bar{c}N_{\pm},
\]
and
\[
{\bf J}_{0,rel} \equiv \frac{1}{P_5} \Big( [{\bf K}_{-} \times {\bf P}_{+}] + [{\bf K}_{+} \times {\bf P}_{-}] \Big)
\]
\[
- \Big( \frac{P_{5,-}}{P_5} {\bf J}_{0,+} + \frac{P_{5,+}}{P_5} {\bf J}_{0,-} \Big)
\]
can be interpreted as an orbital angular momentum related to the relative motion of particles and antiparticles.
The states that only contain particle and antiparticle pairs belong to the zero inertial mass subspace. For those states the Galilean covariant spin operator can not be defined.

The field ${\psi}_{-}({\bf x},t)$ obeys the equation
\[
\left({\rm i}
\gamma^{\bar{\mu}}\partial_{\bar{\mu}}-kI_{-}\right){\psi}_{-}({\bf x},t)=0
\]
and can be quantized in the same way as the field ${\psi}_{+}({\bf x},t)$.  Let $b^{(r)}({\bf p})$ and $b^{{\dagger} (r)}({\bf p})$ be annihilation and creation operators of antiparticles and $|0;- \rangle$ be the vacuum state for negative mass sector:
\[
b^{(r)}({\bf p}) |0;- \rangle = 0 \quad {\rm for} \quad {\rm all} \quad {\rm r}\quad {\rm and} \quad {\bf p}.
\]
One-antiparticle states are defined as
\begin{equation}
|r;{\bf p};- \rangle \equiv b^{{\dagger} (r)}({\bf p}) |0;- \rangle
\end{equation}
being characterized by the mass $(-m)$, the momentum ${\bf p}$, the energy $(-E_p)$ and the spin polarization $r$. The total vacuum state that contains neither particles  nor antiparticles is
\[
|0\rangle = |0;+ \rangle \otimes |0;- \rangle.
\]

We can connect one-particle and one-antiparticle states by making use of the transition operators
\[
T_{-+}(r,s) \equiv \int d^3p\; b^{{\dagger}(r)}({\bf p}) a^{(s)}({\bf p})
\]
and
\[
T_{+-}(r,s) \equiv \int d^3p\; a^{{\dagger}(r)}({\bf p}) b^{(s)}({\bf p})
\]
with
\[
T^{\dagger}_{+-}(r,s) = T_{-+}(s,r).
\]
The operator $T_{-+}(r,s)$ transforms an one-particle state with the spin polarization $s$ to an one-antiparticle state with the spin polarization $r$, while $T_{+-}(r,s)$ performs an inverse transformation:
\begin{eqnarray}
T_{-+}(r,s) |s;{\bf p};+ \rangle \otimes |0;- \rangle & = & |0;+ \rangle \otimes |r;{\bf p};- \rangle, \nonumber \\
T_{+-}(s,r) |0;+ \rangle \otimes |r;{\bf p};- \rangle & = & |s;{\bf p};+ \rangle \otimes |0;- \rangle.\nonumber
\end{eqnarray}
In the transition performed by $T_{-+}(r,s)$ the mass of state decreases by $2m$, and in the transition performed by $T_{+-}(r,s)$ it increases by $2m$. This is reflected in the following commutation relations
\begin{eqnarray}
{[T_{+-}(r,s), M]}_{-} & = & -2m T_{+-}(r,s), \nonumber \\
{[T_{-+}(r,s), M]}_{-} & = & 2m T_{-+}(r,s). \nonumber
\end{eqnarray}
On one-particle (-antiparticle) states,
\[
J_{0,rel}^a |s;{\bf p};+ \rangle \otimes |0;- \rangle =
J_{0,rel}^a |0;+ \rangle \otimes |r;{\bf p};- \rangle = 0
\]
and
\begin{eqnarray}
S^a |s;{\bf p};+ \rangle \otimes |0;- \rangle & = & S_{+}^a |s;{\bf p};+ \rangle \otimes |0;- \rangle ,\nonumber \\
S^a |0;+ \rangle \otimes |r;{\bf p};- \rangle & = & S_{-}^a |0;+ \rangle \otimes |r;{\bf p};- \rangle, \nonumber
\end{eqnarray}
so that the operators $T_{+-}(r,s)$ and $T_{-+}(r,s)$ commute with $(S^a)^2$,
\begin{eqnarray}
{[ T_{+-}(r,s), (S^a)^2]}_{-}\; |0;+ \rangle \otimes |r;{\bf p};- \rangle & = & 0, \nonumber \\
{[ T_{-+}(r,s), (S^a)^2]}_{-}\; |s;{\bf p};+ \rangle \otimes |0;- \rangle & = & 0, \nonumber
\end{eqnarray}
indicating that the magnitude of spin does not change in these transitions.

The transition operators can be used to connect many-particle states, too. Since the mass gap between the positive and negative mass sectors is $2m$, the mass difference between such many-particle states should be an integer multiple of $2m$. By applying $a^{{\dagger}(r)}({\bf p_1}) a^{{\dagger}(s)}({\bf p_2}) ... a^{{\dagger}(t)}({\bf p_{N_{+}}})$ and
$b^{{\dagger}(\bar{r})}({\bf q_1}) b^{{\dagger}(\bar{s})}({\bf q_2}) ... b^{{\dagger}(\bar{t})}({\bf q_{N_{-}}})$, for instance, on the total vacuum state, we create a many-particle state with $N_{+}$-particles and $N_{-}$-antiparticles. The operator $T_{-+}(r,s)$ transforms this state to a state with $(N_{+}-1)$-particles and $(N_{-}+1)$-antiparticles, and the operator $T_{+-}(r,s)$ transforms it to a state with $(N_{+}+1)$-particles and $(N_{-}-1)$-antiparticles.
In transitions between many-particle states, the magnitude of spin is not in general conserved.

\section{Discussion}

1. In $(4+1)$ dimensions, the characteristics of spin of the Galilean covariant Dirac field are shown to be similar to those of the Lorentz covariant one in $(3+1)$ dimensions. The Galilean covariant and Dirac spin operators are introduced in the same way as in the relativistic case, the Galilean covariant spin operator being invariant under time translations. The Dirac spin is not conserved in time, and this is related to an extra symmetry with respect to non-Galilean transformations. These transformations include boosts that connect a Galilean reference frame to non-Galilean ones in which the system has the same energy but different values of the inertial and rest masses. It is shown that the non-Galilean boost generator has a nonvanishing spin part and its spin density is responsible for the nonconservation of Dirac spin.

Reduction to $(3+1)$ dimensions eliminates the extra symmetry and makes boosts to non-Galilean reference frames impossible. The Dirac spin of the resulting nonrelativistic Fermi field is conserved. On one-particle states the Galilean covariant and Dirac spin operators coincide and can be equally used in the construction of spin states.

2. The inertial mass of a nonrelativistic particle is an invariant quantity, while its rest mass can be in principle changed. Two regimes are possible: (i) the inertial and rest masses are equal; in this case the Galilean theory can be obtained as a limit of a Lorentz covariant one, and the parameter $\bar{c}$ can be identified with the speed of light $c$; (ii) the rest mass differs from the inertial one, then there is no direct relation between the Galilean and Lorentz covariant theories, and the parameter $\bar{c}$ differs from $c$. Whether transitions between these two regimes are allowed or not depends on a choice of interactions. If we could design interactions that change the rest mass, this would allow the system to undergo such transitions. If in these transitions the rest energy is conserved, then $mc^2=m_0{\bar{c}}^2$, and this gives us the following relation between the parameters $\bar{c}$ and $c$:
\[
\bar{c} = c{\Big( 1 + \dfrac{{\Delta}m}{m} \Big)}^{-1/2},
\]
where ${\Delta}m=m_0-m$ is the difference between the rest and inertial masses.

\acknowledgments

The author thanks F. Khanna and M. de Montigny for discussions.

\end{document}